\documentclass[aps,prx,reprint,superscriptaddress,nofootinbib]{revtex4-2}

\usepackage{amsmath,amssymb,amsthm,mathtools}
\usepackage{bm}
\usepackage{graphicx}
\usepackage{hyperref}
\usepackage{cleveref}
\usepackage{xcolor}
\usepackage{booktabs}
\usepackage{algorithm}
\usepackage{algpseudocode}

\hypersetup{
    colorlinks=true,
    linkcolor=blue,
    citecolor=blue,
    urlcolor=blue
}

\begin{document}

\title{Planted-solution SAT and Ising benchmarks from integer factorization}

\author{Itay Hen}
\affiliation{Information Sciences Institute, University of Southern California, Marina del Rey, California 90292, USA}
\affiliation{Department of Physics and Astronomy, University of Southern California, Los Angeles, California 90089, USA}

\begin{abstract}
\noindent We present a family of planted-solution benchmark instances for
satisfiability (SAT) solvers and Ising optimization derived from
integer factorization.  Given two primes $p$ and $q$, the
construction encodes the arithmetic constraints of $N = p \times q$
as a conjunctive normal form (CNF) formula whose satisfying
assignments correspond to valid factorizations of~$N$.  The known
pair $(p,q)$ serves as a built-in ground truth, enabling
unambiguous verification of solver output.  We show that for two $d$-bit
primes the total number of carry contractions is on the order of $d^4$.  
Empirical benchmarks with SAT solvers show
that median runtime grows exponentially in the bit-length of the factors over the range tested.
The construction provides a scalable,
structured, and verifiable benchmark family controlled by a single
parameter, accompanied by open-source generation software.
\end{abstract}

\maketitle
%% ====================================================================
\section{Introduction}
\label{sec:intro}
%% ====================================================================

Benchmarking satisfiability and optimization solvers requires
instance families that simultaneously exhibit realistic structure,
systematic scalability, and verifiable ground truth~\cite{Achlioptas2004RandomKSAT}.
These requirements are difficult to satisfy within a single
framework. Random ensembles---such as uniform random $k$-SAT---can
be generated at arbitrary sizes and are hard near the satisfiability
threshold, but they lack a planted solution against which solver
output can be validated~\cite{Feldman2013PlantedKSAT,Krzakala2012QuietPlanting}. Conversely, crafted instances from
competition libraries provide structure but typically do not offer
a controlled, single-parameter scaling of difficulty~\cite{Balint2015SATCompetition}.

Planted-solution constructions address this gap by engineering
instances whose optimal solution is known by design.
This paradigm has been explored in several directions.
In the context of Ising optimization, frustration-based planted
spin-glass instances~\cite{Hen2015} and equation planting~\cite{Hen2019}
provide scalable benchmarks with known ground states.
Structured planted XORSAT ensembles have been used to compare
classical and quantum heuristic optimizers~\cite{Kowalsky2022}.
In the SAT literature, planted $k$-SAT constructions offer a way
to control hardness while retaining a known satisfying assignment
\cite{Jia2005,Achlioptas2005}.
However, most existing planted ensembles are based on random disorder
or algebraic planting, and therefore do not capture the deterministic,
long-range structure characteristic of many real computational
problems.

In this work, we introduce a qualitatively different class of
planted benchmark instances derived from integer factorization.
Given two primes $p$ and $q$, we encode the arithmetic constraints
of the product $N = pq$ as a CNF formula whose satisfying
assignments correspond to valid factorizations of~$N$, with the
pair $(p,q)$ serving as the planted solution.
The construction is controlled by a single parameter---the bit-length
$d$ of each factor---and produces instances whose structure is
inherited directly from the binary multiplication circuit.

The key distinguishing feature of this construction is the presence
of carry-induced long-range correlations.
Each local contraction in the multiplication circuit generates
carries that propagate across columns, producing a cascade that
couples variables separated by distances of order $d^2$.
This mechanism leads to a quartic growth in instance size and
induces a nontrivial interaction structure that is fundamentally
different from both random SAT ensembles and previously studied
planted constructions.

The idea of encoding factorization as a constraint satisfaction
problem is classical; early work noted that multiplication circuits
can be translated into SAT formulas for the purpose of factoring
integers~\cite{factoring_as_sat,CookMitchell1997,MassacciMarraro2000}.
Our goal is different.
We do not view the SAT encoding as a route to solving factorization,
but rather as a benchmark generator: a single-parameter family
of instances with (i)~exactly analyzable size scaling,
(ii)~a built-in planted solution for verification,
(iii)~a deterministic and interpretable structure arising from
arithmetic carry propagation, and (iv)~a direct compilation into
both CNF and quadratic Ising form.

These features place the present construction in a complementary
regime to existing benchmark families. Unlike random or
frustration-based planted instances, the structure here is not
derived from disorder but from arithmetic constraints with
intrinsic long-range dependencies. This makes the resulting
instances a useful testbed for probing solver behavior on
structured, non-random problems.

The remainder of this paper is organized as follows.
Section~\ref{sec:construction} describes the three-stage pipeline:
clause generation, Boolean preprocessing, and CNF output. We then derive exact closed-form expressions for
all pre-reduction instance-size quantities and prove that the
leading-order scaling is~$\Theta(d^4)$ in Sec.~\ref{sec:scaling}. Section~\ref{sec:benchmark} reports solver benchmarks characterizing
the empirical difficulty of the instances.
Section~\ref{sec:ising} describes the Ising compilation, and Sec.~\ref{sec:summary} summarizes the results and discusses
future directions.

%% ====================================================================
\section{Construction}
\label{sec:construction}
%% ====================================================================

We select two prime numbers $p$ and $q$, represented in binary as
\begin{equation}
p = \sum_{i=0}^{n_p-1} p_i \, 2^i,
\qquad
q = \sum_{j=0}^{n_q-1} q_j \, 2^j,
\end{equation}
with $p_i, q_j \in \{0,1\}$ and $n_p, n_q$ denoting the respective
bit-lengths.  Their product
\begin{equation}
N = pq = \sum_{k=0}^{n_p+n_q-2} N_k \, 2^k
\end{equation}
is computed and its binary digits $N_k$ are treated as known
constants.  The goal is to construct a CNF formula $\phi$ over
Boolean variables representing the unknown bits of $p$ and $q$
(together with auxiliary variables) such that a satisfying assignment
of $\phi$ encodes a valid factorization of~$N$.

The pipeline consists of three stages: (i)~clause generation from
binary multiplication, (ii)~Boolean preprocessing via iterative
logical reduction, and (iii)~conversion to DIMACS CNF.

%% ------------------------------------------------------------------
\subsection{Clause generation from binary multiplication}
\label{subsec:clause_generation}

Binary multiplication of $p$ and $q$ proceeds by the standard
shift-and-add algorithm.  Each partial product $a_{ij} = p_i \wedge
q_j$ is a Boolean AND operation, producing an auxiliary variable
that is placed in column $k = i + j$ of the multiplication table.
Column~$k$ thus receives
\begin{equation}
\mathrm{pp}_k = \min(k{+}1,\; n_p,\; n_q,\; n_p{+}n_q{-}1{-}k)
\label{eq:pp_k}
\end{equation}
partial products.

When a column contains more than one entry, the entries are
contracted pairwise using the half-adder decomposition: for two
entries $x$ and $y$ in the same column,
\begin{equation}
\mathrm{sum} = x \oplus y, \qquad \mathrm{carry} = x \wedge y,
\label{eq:half_adder}
\end{equation}
where the sum remains in the current column and the carry is
promoted to the next column.  Each contraction introduces two new
auxiliary variables (one for the sum, one for the carry) and two new
constraints (one XOR clause and one AND clause).  The contractions
are applied iteratively until each column contains a single entry.
After full contraction, the single remaining entry in each column is
matched to the known bit $N_k$: bits of $N$ that are~$1$ force the
corresponding variable to \textsc{True}, and bits that are~$0$ force
it to \textsc{False}.  These assignments form the set of
pinning constraints.

The output of this stage is a system of three types of constraints: (i) AND clauses: $c = a \wedge b$, encoding partial
  products and carries. (ii) XOR clauses: $c = a \oplus b$, encoding column
  sums. (iii) Pinning constraints: variables fixed to
  \textsc{True} or \textsc{False} by matching to the known bits
  of~$N$.
%% ------------------------------------------------------------------
\subsection{Boolean preprocessing}
\label{subsec:preprocessing}

Before conversion to CNF, the constraint system undergoes iterative
logical simplification to reduce the number of
variables and clauses.  The reduction loop cycles until convergence,
applying four operations at each iteration: (i) Pin propagation: Variables fixed by the bits of $N$
  or by prior inference are substituted throughout all clauses.
(ii) AND simplification: Case analysis on AND clauses
  performs constant folding (e.g., $0 \wedge x = 0$), variable
  identification (e.g., $x \wedge x = x$), and complementary-pair
  detection ($x \wedge \bar{x} = 0$).
(iii) XOR simplification: Analogous case analysis on XOR
  clauses identifies equal variables ($x \oplus x = 0$ implies
  output is pinned), anti-equal variables ($x \oplus \bar{x} = 1$),
  and constant folding.
(iv) Cross-clause inference: Shared variables between AND
  and XOR clauses enable further deductions (e.g., if an AND
  clause's output appears in a XOR clause that pins it to a known
  value, the AND inputs become constrained).

Throughout this process, variables identified as equal (or equal up
to negation) are merged into equivalence classes maintained by a
union-find data structure.  All subsequent operations use the
canonical representative of each class, reducing the effective
variable count.  Only the residual clauses---those that
cannot be resolved by logical deduction---survive to the output
stage. A worked example is provided in the appendix.

Figure~\ref{fig:preprocessing} shows the effect of preprocessing on
instance size as a function of the bit-length~$d$.  At small~$d$,
preprocessing eliminates most of the instance: for $d = 4$, over
80\%\ of the raw CNF clauses are removed.  As $d$ grows, the
fraction of clauses eliminated decreases, and the post-reduction
instance size approaches the pre-reduction $d^4$ scaling.  This
trend reflects the fact that the number of variables directly
constrained by the pinning conditions ($d^2$ pins) grows more
slowly than the total instance size ($\sim d^4$), leaving an
increasingly large ``hard core'' of residual constraints.

\begin{figure}[t]
\centering
\includegraphics[width=\columnwidth]{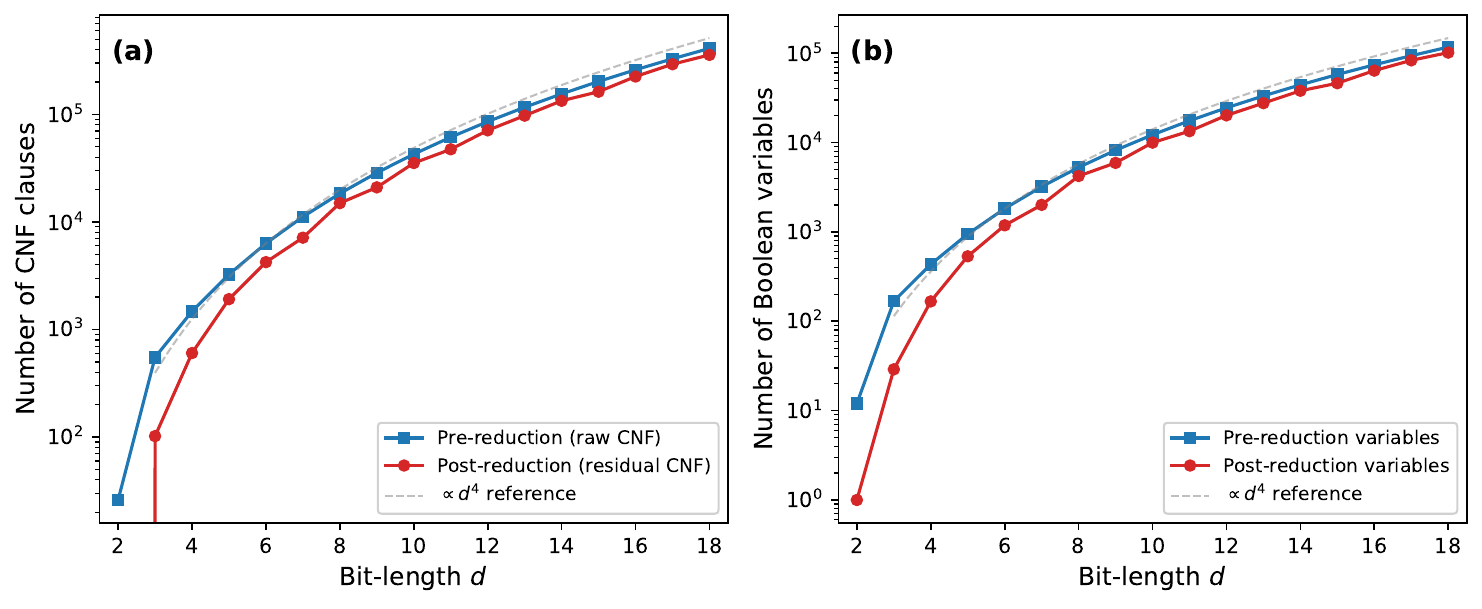}
\caption{Effect of Boolean preprocessing on instance size for
$d = 2, \ldots, 19$ (symmetric case $n_p = n_q = d$, one random
prime pair per~$d$).
(a)~Number of CNF clauses before (blue squares) and after (red
circles) preprocessing, with a $d^4$ reference line (dashed gray).
(b)~Number of Boolean variables before and after preprocessing.
Both pre-reduction counts follow the exact $d^4$ scaling of
Table~\ref{tab:formulas}.  The gap between pre- and post-reduction
narrows with increasing~$d$, indicating that preprocessing becomes
proportionally less effective for larger instances.}
\label{fig:preprocessing}
\end{figure}

%% ------------------------------------------------------------------
\subsection{Conversion to DIMACS CNF}
\label{subsec:cnf_conversion}

The residual AND and XOR clauses are converted to conjunctive normal
form using standard encodings.
An AND clause $c = a \wedge b$ is encoded as three CNF clauses:
\begin{equation}
(\bar{a} \vee \bar{b} \vee c), \quad
(a \vee \bar{c}), \quad
(b \vee \bar{c}).
\label{eq:and_cnf}
\end{equation}
These enforce the AND relation exactly: $c = 1$ if and only if
$a = b = 1$.

A XOR clause $c = a \oplus b$ is encoded as four CNF clauses:
\begin{equation}
(\bar{c} \vee \bar{a} \vee \bar{b}), \quad
(\bar{c} \vee a \vee b), \quad
(c \vee \bar{a} \vee b), \quad
(c \vee a \vee \bar{b}).
\label{eq:xor_cnf}
\end{equation}
These enforce the XOR relation exactly: $c = 1$ if and only if
exactly one of $a, b$ is~$1$.

Variables that were pinned during preprocessing are treated as
constants: any clause containing a true literal is satisfied and
omitted, and false literals are dropped from clauses.  The surviving
variables are reindexed contiguously starting from~$1$.
The final output is a standard DIMACS CNF file that can be fed
directly to any SAT solver.
The pre-reduction instance size is $\mathcal{O}(d^4)$, dominated
by the contraction count derived in Sec.~\ref{sec:scaling};
empirically, the full pipeline (clause generation, preprocessing,
and CNF/Ising export) scales accordingly over the range tested.

%% ====================================================================
\section{Instance-Size Scaling}
\label{sec:scaling}
%% ====================================================================

We now derive exact closed-form expressions for the number of
variables and clauses produced by the pipeline before
preprocessing.  We focus on the symmetric case $n_p = n_q = d$; the
general case $n_p \neq n_q$ is discussed at the end of this
section.

%% ------------------------------------------------------------------
\subsection{Column-entry recurrence}

Let $m_k$ denote the total number of entries (partial products plus
incoming carries) in column~$k$ at the time it is processed.  Since
a column with $m_k \geq 2$ entries requires $m_k - 1$ pairwise
contractions, each producing one carry for the next column, the
entries obey the recurrence
\begin{equation}
m_{k+1} = \mathrm{pp}_{k+1} + \max(m_k - 1,\, 0),
\qquad m_0 = 1.
\label{eq:recurrence}
\end{equation}

%% ------------------------------------------------------------------
\subsection{Exact solution in three phases}

The column sequence decomposes into three phases with distinct
behavior. (i) Ascending phase ($k = 0, \ldots, d{-}1$):
In this range $\mathrm{pp}_k = k{+}1$.  Solving the
recurrence~\eqref{eq:recurrence} with $m_0 = 1$ gives
\begin{equation}
m_k = 1 + \frac{k(k{+}1)}{2}\,, \qquad k = 0, \ldots, d{-}1.
\label{eq:mk_asc}
\end{equation}
This is verified by substitution: $m_{k+1} = (k{+}2) + m_k - 1 =
k + 1 + m_k$, consistent with~\eqref{eq:mk_asc}.  The entries
grow quadratically, reaching $m_{d-1} = 1 + d(d{-}1)/2$ at the end
of the ascending phase.
(ii) Descending phase ($k = d, \ldots, 2d{-}2$):
Here $\mathrm{pp}_k = 2d{-}1{-}k$.  Writing $j = k - d$, the
recurrence yields
\begin{equation}
m_{d+j} = 1 + \frac{d(d{-}1)}{2}
          + (j{+}1)(d{-}2) - \frac{j(j{+}1)}{2}\,.
\label{eq:mk_desc}
\end{equation}
The entries continue to grow but with decelerating increments,
reaching a peak at $k = 2d{-}2$:
\begin{equation}
m_{2d-2} = 1 + (d{-}1)^2 = d^2 - 2d + 2.
\label{eq:peak}
\end{equation}
(iii) Tail phase ($k \geq 2d{-}1$): 
Beyond column $2d{-}2$ there are no more partial products.  Each
column receives $m_{k-1} - 1$ carries, so $m_k = m_{k-1} - 1$: the
entries decrease by one per column until $m_k = 1$.  The tail
contains $(d{-}1)^2$ columns, and the last active column is
\begin{equation}
k_{\max} = 2d - 2 + (d{-}1)^2 = d^2 - 1.
\label{eq:kmax}
\end{equation}
The total number of active columns is therefore $d^2$.

\begin{figure}[t]
\centering
\includegraphics[width=\columnwidth]{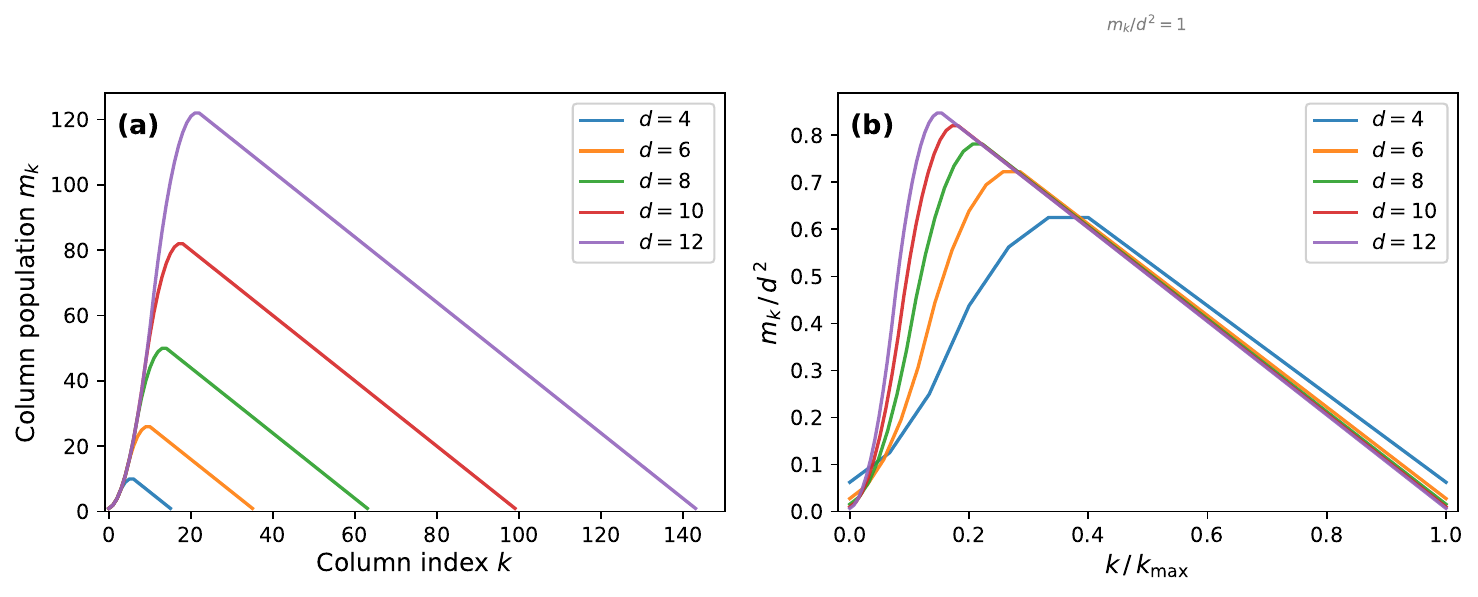}
\caption{Column-population profiles $m_k$ for
$d = 4, 6, 8, 10, 12$.
(a)~Raw column population $m_k$ versus column index~$k$.
The three phases---ascending, descending, and tail---produce a
characteristic asymmetric peak whose height scales as $d^2$
[Eq.~\eqref{eq:peak}] and whose support spans $d^2$ columns.
(b)~Rescaled profiles $m_k / d^2$ versus $k / k_{\max}$.
The tail phases collapse onto a single line, and the overall shape
converges to a universal curve as $d$ increases.}
\label{fig:mk_profiles}
\end{figure}

Figure~\ref{fig:mk_profiles} illustrates the column-population
sequence $m_k$ for several values of~$d$.  The three-phase structure
is clearly visible: a rapid quadratic ramp during the ascending
phase, a broad peak near $k = 2d{-}2$, and a long linear decay
during the tail phase. When rescaled as $m_k / d^2$ versus
$k / k_{\max}$, the curves collapse onto a universal profile that
converges toward a limiting shape as $d \to \infty$, with the
rescaled peak approaching unity.

%% ------------------------------------------------------------------
\subsection{Total contractions}

The total number of pairwise contractions across all columns,
$C(d) = \sum_{k} \max(m_k - 1, 0)$, is obtained by summing the
three phases. 
For two $d$-bit factors, the contributions from the three phases are:
\begin{align}
C_1 &= \sum_{k=0}^{d-1}\frac{k(k{+}1)}{2}
     = \frac{d(d{-}1)(d{+}1)}{6}\,,
     \label{eq:C1} \\[3pt]
C_3 &= \sum_{t=1}^{(d-1)^2 - 1} t
     = \frac{\bigl[(d{-}1)^2{-}1\bigr](d{-}1)^2}{2}\,.
     \label{eq:C3}
\end{align}
The descending-phase sum $C_2 = \sum_{j=0}^{d-2}(m_{d+j} - 1)$ is
evaluated using~\eqref{eq:mk_desc}.  Adding $C_1 + C_2 + C_3$ and
simplifying yields
\begin{equation}
\;C(d) = \frac{d^2(d{-}1)^2}{2}\,,\;
\label{eq:Cd}
\end{equation}
the total number of half-adder contractions.

%We have verified the formula numerically for all $d = 2, \ldots, 512$.

%% ------------------------------------------------------------------
\subsection{Pre-reduction instance sizes}

Each contraction produces one AND clause (carry) and one XOR clause,
and introduces two new auxiliary variables.  Together with the $d^2$
AND clauses from partial products and $d^2$ pinning constraints, the
complete pre-reduction counts are:

\begin{table}[h]
\centering
\caption{Pre-reduction instance sizes for two $d$-bit factors.  All
expressions are exact.}
\label{tab:formulas}
\begin{tabular}{lll}
\toprule
Quantity & Exact expression & Leading order \\
\midrule
Partial products & $d^2$ & $d^2$ \\[2pt]
Contractions $C(d)$ & $d^2(d{-}1)^2/2$ & $d^4/2$ \\[2pt]
Boolean variables
  & $2d + d^2 + d^2(d{-}1)^2$
  & $d^4$ \\[2pt]
AND clauses
  & $d^2 + d^2(d{-}1)^2/2$
  & $d^4/2$ \\[2pt]
XOR clauses
  & $d^2(d{-}1)^2/2$
  & $d^4/2$ \\[2pt]
Pinning constraints
  & $d^2$
  & $d^2$ \\[2pt]
Total constraints
  & $2d^2 + d^2(d{-}1)^2$
  & $d^4$ \\[2pt]
Active columns
  & $d^2$ & $d^2$ \\[2pt]
Peak column entries
  & $d^2 - 2d + 2$
  & $d^2$ \\
\bottomrule
\end{tabular}
\end{table}

The $d^4$ scaling is a
direct consequence of carry cascading.  Each column contraction
injects a carry into the next column, which in turn requires
additional contractions that generate further carries.  This
positive-feedback loop causes the column-entry count~$m_k$ to grow
quadratically in~$k$ [Eq.~\eqref{eq:mk_asc}], and summing a
quadratic over $\sim d^2$ active columns yields a quartic total.
Physically, carry cascading is the arithmetic analogue of long-range
correlation propagation: a single bit flip in a low-order column can
alter carries that propagate all the way to column $d^2 - 1$.

In the DIMACS CNF output, each AND clause becomes 3~CNF clauses
[Eq.~\eqref{eq:and_cnf}] and each XOR clause becomes 4~CNF clauses
[Eq.~\eqref{eq:xor_cnf}].  The total number of CNF clauses before
preprocessing is therefore
\begin{equation}
\text{CNF clauses} = 3\Bigl(d^2 + \tfrac{d^2(d{-}1)^2}{2}\Bigr)
                   + 4 \cdot \tfrac{d^2(d{-}1)^2}{2}
                   \sim \tfrac{7}{2}\,d^4\,.
\label{eq:cnf_total}
\end{equation}

%% ====================================================================
\section{Benchmark Study}
\label{sec:benchmark}
%% ====================================================================

To assess the computational difficulty of the planted factorization
instances, we carried out a systematic solver benchmark.  The
central result is that median solver runtime is consistent with
exponential scaling in the bit-length~$d$ of the factors over the
range tested, with both solvers exhibiting a remarkably steady
doubling of runtime per additional bit.

For bit-lengths $d$ ranging from 8 to~27, we generated random
prime pairs $(p,q)$ with $\max(n_p, n_q) = d$ and constructed,
preprocessed, and exported each instance in DIMACS CNF format as
described in Sec.~\ref{sec:construction}.

We evaluated two state-of-the-art conflict-driven clause-learning
(CDCL) SAT solvers: Kissat~3.0~\cite{Kissat} and
CaDiCaL~1.5~\cite{CaDiCaL}, both run with default parameters in
single-threaded mode on a machine equipped with an Intel Xeon CPU
@ 3.2~GHz and 64~GB RAM.  The planted solution $(p,q)$ was
verified against the solver output in every case.

\begin{figure}[t]
\centering
\includegraphics[width=\columnwidth]{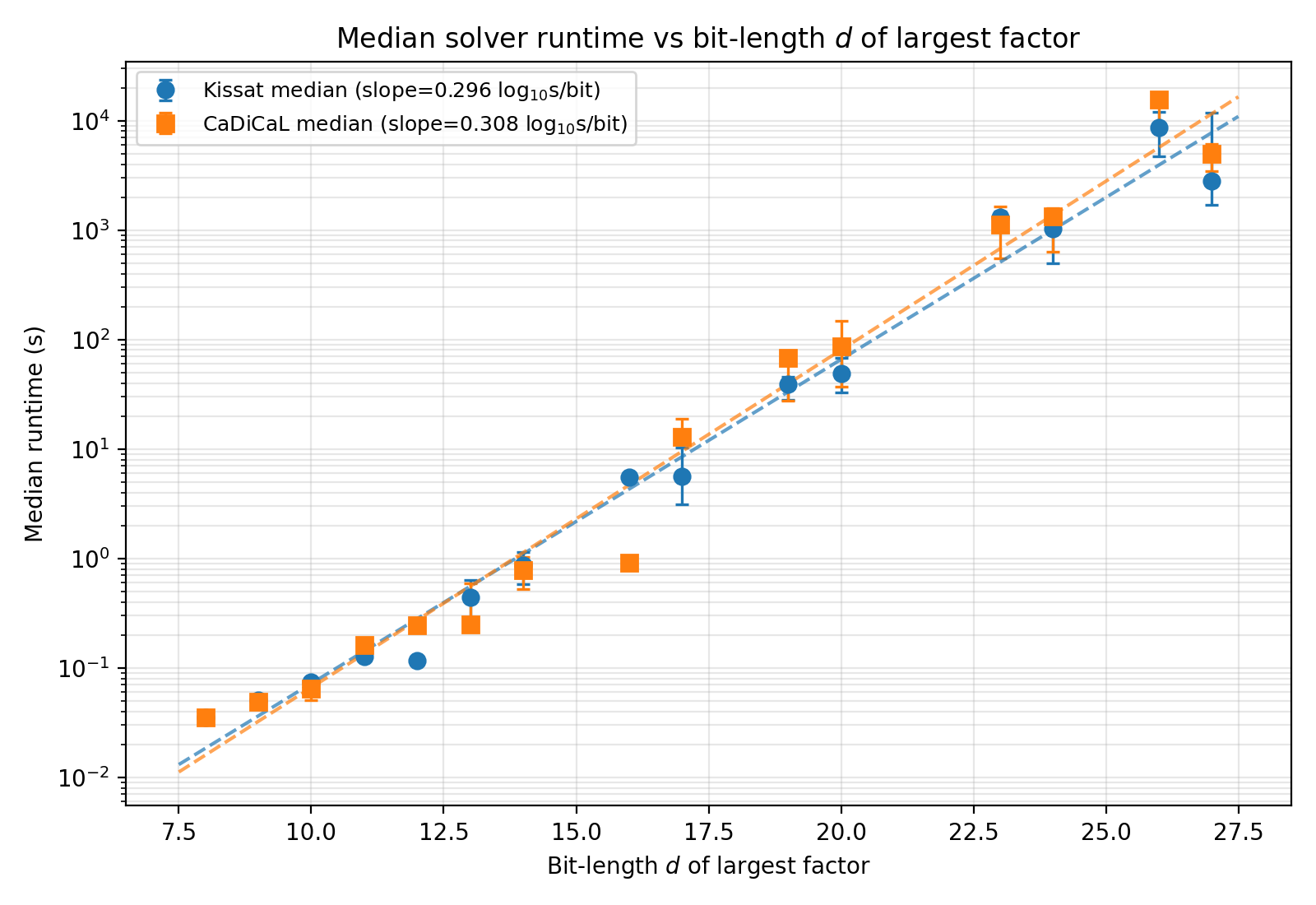}
\caption{Median solver runtime versus the bit-length~$d$ of the
largest factor in each instance.  Dashed lines show log-linear fits
[Eq.~\eqref{eq:loglinear_fit}]; the fitted slopes are
$\alpha = 0.296$ (Kissat) and $\alpha = 0.308$ (CaDiCaL),
corresponding to an approximately twofold increase in runtime per
additional bit.}
\label{fig:runtime_scaling}
\end{figure}

Figure~\ref{fig:runtime_scaling} shows the median solver runtime as
a function of the bit-length~$d$, plotted on a logarithmic vertical
axis.  The data for both solvers fall on approximately straight
lines, indicating that the median runtime~$T$ grows exponentially
with~$d$:
\begin{equation}
\log_{10} T \;\approx\; \alpha \, d + \mathrm{const.}
\label{eq:loglinear_fit}
\end{equation}
A least-squares fit yields slopes $\alpha = 0.296$ for Kissat and
$\alpha = 0.308$ for CaDiCaL. Since
$10^{\alpha} \approx 2$, each additional bit roughly doubles the
median runtime.  Equivalently, the scaling can be written as
\begin{equation}
T_{\mathrm{median}} \;\sim\; 2^{\,\beta\,d}\,,
\label{eq:exp_scaling}
\end{equation}
with $\beta = \alpha / \log_{10} 2 \approx 0.98$ (Kissat) and
$\approx 1.02$ (CaDiCaL)---both remarkably close to unity,
corresponding to a doubling of runtime per additional bit.
Since the number of bits is $d \approx \log_2 N$, this represents
exponential scaling in the input size~$d$.
The modest instance sizes tested here ($d \leq 27$, i.e.,
$N \lesssim 10^{16}$) are still in the regime where CDCL heuristics
and unit propagation resolve much of the search.

The two solvers track each other closely across the entire range,
with Kissat holding a small but consistent speed advantage.  The
near-identical slopes suggest that the growth rate is governed by
structural features of the multiplication circuit---particularly
the long-range carry correlations analyzed in
Sec.~\ref{sec:scaling}---rather than by solver-specific heuristics.

The empirical data indicate that solver difficulty grows rapidly and
controllably with the bit-length.  At $d = 27$ (corresponding to
$N \sim 10^{16}$), median runtimes already reach $\sim\!10^4$~s,
suggesting that instances with $d \geq 35$--$40$ will provide a
meaningful stress test for modern CDCL solvers.  The quartic growth
of the CNF instance size (Sec.~\ref{sec:scaling}) compounds this:
both the number of clauses and the empirical search effort scale
rapidly with~$d$, making the construction a practical source of
increasingly challenging benchmarks.

The close agreement between the two solvers also suggests that
performance comparisons on these instances are unlikely to be
artifacts of solver-specific tuning, since both solvers encounter
the same underlying constraint structure.

%% ====================================================================
\section{Ising Compilation}
\label{sec:ising}
%% ====================================================================

For applications to classical and quantum optimization, the residual
clause system produced by the preprocessing pipeline of
Sec.~\ref{sec:construction} can be compiled into a quadratic Ising
Hamiltonian
\begin{equation}
H(\mathbf{s}) = E_0 + \sum_i h_i\,s_i + \sum_{i<j} J_{ij}\,s_i s_j\,,
\label{eq:ising_H}
\end{equation}
with $s_i \in \{+1, -1\}$, whose ground state encodes the planted
factorization.  The compilation proceeds in three steps: (i)~each
Boolean variable $b_i$ is mapped to an Ising spin via
$s_i = 2b_i - 1$ (equivalently, $b_i = (1 + s_i)/2$); (ii)~each
residual AND and XOR clause is replaced by an energy gadget---a
low-degree polynomial in the spins---that evaluates to zero on
satisfying assignments and to a strictly positive value otherwise;
(iii)~the gadgets are summed, duplicate spin-pair couplings are
merged by adding their weights, and all constant terms are absorbed
into the offset~$E_0$.

\subsection{Energy gadgets}
\label{subsec:gadgets}

Each constraint type admits a natural exact
energy function that is quadratic in the Ising spins and
encodes the logical relation exactly.  We derive both gadgets by
substituting $b_i = (1 + s_i)/2$ into a penalty function that is
zero on satisfying Boolean assignments and strictly positive
otherwise.

The AND constraint $c = a \wedge b$ is satisfied by exactly four of
the eight Boolean assignments $(a, b, c) \in \{0,1\}^3$:
$(0,0,0)$, $(0,1,0)$, $(1,0,0)$, and $(1,1,1)$.
A penalty that vanishes on these assignments and equals~1 on
the remaining four is
$P_{\wedge}(a,b,c) = ab - c$
(easily verified by enumeration).  However, this is not
non-negative.  The standard non-negative quadratic penalty is
\begin{equation}
P_{\wedge}(a,b,c) = ab - 2ac - 2bc + 3c\,,
\label{eq:and_bool_penalty}
\end{equation}
which takes the value~$0$ on satisfying assignments and positive
integer values ($1$, $1$, $1$, $3$) on the four violating ones.
Substituting $a = (1+s_1)/2$, $b = (1+s_2)/2$, $c = (1+s_3)/2$
and simplifying yields the three-spin Ising gadget
\begin{equation}
E_{\wedge}(s_1,s_2,s_3) = 3 - s_1 - s_2 + s_1 s_2
  + 2 s_3 - 2 s_1 s_3 - 2 s_2 s_3\,,
\label{eq:and_gadget}
\end{equation}
which evaluates to~$0$ on the four satisfying spin configurations and
to~$4$, $4$, $4$, or~$12$ on the four violating ones.  The spectral
gap $\Delta_{\wedge} = 4$ ensures that any single violated AND
constraint incurs a penalty of at least~4.

The XOR constraint $c = a \oplus b$ relates three Boolean variables:
$c = 1$ if and only if exactly one of $a$, $b$ is~$1$.
Under the Ising mapping, the XOR relation becomes
$s_3 = -s_1 s_2$,
which is a cubic monomial in the spins and therefore cannot be
represented by a quadratic Hamiltonian without auxiliary variables.
To reduce to quadratic order, we introduce one auxiliary spin $s_a$
whose planted value encodes the product $s_1 s_2$.  Specifically,
the Boolean penalty
\begin{equation}
P_{\oplus}(a,b,c) = a + b + c - 2ab - 2ac - 2bc + 4abc
\label{eq:xor_bool_penalty}
\end{equation}
vanishes on the four satisfying assignments of $c = a \oplus b$ and
equals~$1$ on the four violating ones.  The quartic term $abc$ is
eliminated by introducing an auxiliary bit $d$ and replacing $abc$
with a quadratic surrogate that, upon minimization over $d$, recovers
the correct penalty.  The resulting four-spin Ising gadget is
\begin{align}
E_{\oplus}(s_1,s_2,s_3,s_a)
&= 4 - s_1 - s_2 + s_1 s_2 + s_3 - s_1 s_3 - s_2 s_3 \nonumber\\
&\quad + 2\,s_a - 2\,s_1 s_a - 2\,s_2 s_a + 2\,s_3 s_a\,,
\label{eq:xor_gadget}
\end{align}
which evaluates to~$0$ when the XOR relation holds (with the correct
auxiliary) and to $\geq 2$ on all violating configurations.  The
planted auxiliary value is
\begin{equation}
s_a^{\star} =
\begin{cases}
+1 & \text{if } s_1 = s_2 = +1\,,\\
-1 & \text{otherwise}\,,
\end{cases}
\label{eq:xor_aux_planted}
\end{equation}
which in Boolean language corresponds to $d^{\star} = a \wedge b$,
i.e., the auxiliary records the AND of the two XOR inputs.  The
spectral gap is $\Delta_{\oplus} = 2$.
Table~\ref{tab:gadgets} collects the key properties of both gadgets.
\begin{table}[h]
\centering
\caption{Properties of the Ising energy gadgets.}
\label{tab:gadgets}
\small
\begin{tabular}{lcccc}
\toprule
Gadget & Physical spins & Auxiliary spins & Couplings & Gap \\
\midrule
AND  & 3 & 0 & 3 & 4 \\
XOR  & 3 & 1 & 6 & 2 \\
\bottomrule
\end{tabular}
\end{table}

\noindent The AND gadget is natively quadratic (no auxiliary spins),
while each XOR gadget introduces one auxiliary spin.  Since the
preprocessing pipeline produces $n_{\wedge}$ residual AND clauses and
$n_{\oplus}$ residual XOR clauses, the Ising model contains
$n_{\mathrm{phys}}$ physical spins (one per surviving Boolean variable)
plus $n_{\oplus}$ auxiliary spins, for a total of
$n_{\mathrm{phys}} + n_{\oplus}$ spins.

\subsection{Hamiltonian assembly}
\label{subsec:assembly}

The full Hamiltonian is obtained by summing the energy contributions
of all gadgets:
\begin{equation}
H(\mathbf{s}) = \sum_{\alpha \in \mathcal{A}} E_{\wedge}^{(\alpha)}
  + \sum_{\beta \in \mathcal{X}} E_{\oplus}^{(\beta)}\,,
\label{eq:H_sum}
\end{equation}
where $\mathcal{A}$ and $\mathcal{X}$ denote the sets of residual
AND and XOR clauses, respectively.  In practice, the assembly
proceeds as follows. (i) Spin indexing: Each surviving Boolean variable after preprocessing is assigned
a unique spin index $i \in \{1, \ldots, n_{\mathrm{phys}}\}$.
For each XOR clause~$\beta$, an auxiliary spin
$s_{a_\beta}$ is added with index
$n_{\mathrm{phys}} + \beta$.
(ii) Gadget expansion: Each AND clause $(s_i, s_j, s_k)$ contributes the local-field
and coupling terms from Eq.~\eqref{eq:and_gadget}, and each XOR
clause $(s_i, s_j, s_k, s_{a_\beta})$ contributes those from
Eq.~\eqref{eq:xor_gadget}.  These contributions are accumulated
into the arrays $\{h_i\}$ and $\{J_{ij}\}$.
(iii) Coupling merging: When multiple gadgets share a spin pair $(i,j)$, the
corresponding couplings $J_{ij}$ are summed.  This is the key
step that generates the interaction graph of the Ising model:
the graph inherits both the local structure of individual
gadgets and the global connectivity induced by variables that
participate in multiple clauses.  Shared variables---especially
those along carry chains---produce dense coupling neighborhoods,
reflecting the long-range correlations discussed in
Sec.~\ref{sec:scaling}.
(iv) Constant absorption: All constant terms from the gadget expansions are collected
into the offset $E_0$.

The planted spin configuration
$\mathbf{s}^{\star}$---comprising the Ising images of the known bits
of $p$ and $q$, all internal auxiliary variables determined by the
clause structure, and the XOR auxiliary spins fixed by
Eq.~\eqref{eq:xor_aux_planted}---achieves $H(\mathbf{s}^{\star}) =
E_0$ by construction, since every gadget evaluates to zero on the
satisfying assignment.  Any spin configuration that violates at least
one clause incurs a penalty of at least
$\min(\Delta_{\wedge}, \Delta_{\oplus}) = 2$ above~$E_0$.
For distinct primes $p \neq q$ with $n_p = n_q$, the factorization
$N = pq$ admits exactly two ordered representations, $(p,q)$ and
$(q,p)$, both encodable within the same variable layout, so the
CNF formula has exactly two satisfying assignments and the Ising
ground state is twofold degenerate.  (In the asymmetric case
$n_p \neq n_q$, the swapped pair does not fit the variable layout,
yielding a unique satisfying assignment.)

\subsection{Instance-size scaling of the Ising model}
\label{subsec:ising_scaling}

We can give exact pre-reduction counts for the Ising model in the
symmetric case $n_p = n_q = d$.  The number of AND clauses is
$n_{\wedge} = d^2 + C(d)$ (partial-product ANDs plus carry ANDs),
and the number of XOR clauses is $n_{\oplus} = C(d) =
d^2(d-1)^2/2$.  Each AND gadget contributes 3 couplings and each XOR
gadget contributes~6, giving an upper bound on the total number of
couplings (before merging) of
\begin{equation}
|\{J_{ij}\}|_{\text{raw}} \leq 3\bigl(d^2 + C(d)\bigr) + 6\,C(d)
= 3d^2 + 9\,C(d) \sim \tfrac{9}{2}\,d^4\,.
\label{eq:coupling_count}
\end{equation}
After merging duplicate pairs, the actual number of distinct couplings
is smaller, but the $\Theta(d^4)$ leading order is preserved.

The total spin count is
\begin{eqnarray}
n_{\text{spins}} &=& n_{\mathrm{phys}} + n_{\oplus}
=\bigl(2d + d^2 + 2C(d)\bigr) + C(d) 
\nonumber \\ 
&=& 
2d + d^2 + 3C(d)
\sim \tfrac{3}{2}\,d^4\,,
\label{eq:spin_count}
\end{eqnarray}
where $n_{\mathrm{phys}} = 2d + d^2 + 2C(d)$ counts the $2d$~input
bits, $d^2$~partial products, and $2C(d)$~sum and carry variables from
contraction.  The XOR auxiliaries add another $C(d)$ spins.

\subsection{Interaction graph structure}
\label{subsec:graph_structure}

The interaction graph $G = (V, E)$ of the Ising Hamiltonian---where
vertices are spins and edges connect pairs with nonzero
$J_{ij}$---inherits a distinctive structure from the multiplication
circuit.  Unlike random spin-glass models with uniform or
Erd\H{o}s--R\'enyi connectivity, the present graph has several
notable features: (i) Heterogeneous degree distribution.
Spins corresponding to carries in the high-population columns
(near $k \approx 2d{-}2$) participate in many gadgets and
therefore have high degree, while input-bit spins and
low-column carries have low degree.  The maximum degree scales
as $\Theta(d^2)$, reflecting the peak column population
[Eq.~\eqref{eq:peak}].
(ii) Long-range edges.
Carry propagation couples spins originating in distant columns
of the multiplication table.  In the Ising graph, this manifests
as edges connecting spins whose column indices differ by
$\mathcal{O}(d^2)$, producing long-range edges reminiscent of
small-world connectivity and absent in local lattice models.
(iii) Hierarchical community structure.
The column-by-column contraction naturally groups spins into
communities (one per column), with inter-community edges created
exclusively by carry promotion.  This hierarchical structure
could be exploited by decomposition-based optimization algorithms.

Several features of the Ising compilation are worth emphasizing.
First, the gadgets in Eqs.~\eqref{eq:and_gadget}
and~\eqref{eq:xor_gadget} have no free penalty parameters:
the coefficients are fixed by the
requirement that the energy vanishes on satisfying assignments and is
strictly positive otherwise, while remaining at most quadratic in the
spins.  This eliminates a common source of arbitrariness in
penalty-based Ising compilations, where the choice of penalty
strengths can influence solver performance and obscure intrinsic
problem difficulty.

Second, the ground-state energy $E_0$ and the planted configuration
$\mathbf{s}^{\star}$ are known analytically, providing an exact
ground-truth energy for benchmarking. The minimum spectral gap above
the ground state is $\Delta = 2$ (set by the XOR gadget), meaning
that any single constraint violation costs at least 2 energy units.
For configurations violating $v$ constraints, the energy satisfies
$H \geq E_0 + 2v$, giving the optimization landscape a structured,
``staircase'' penalty structure.

Finally, the dual CNF/Ising representation of the same problem
instance enables direct cross-platform benchmarking: the identical
factorization problem can be solved by a SAT solver (via the DIMACS
file), a classical Ising optimizer (e.g., simulated annealing,
parallel tempering), or a quantum optimizer (after embedding), with
the planted solution serving as a common ground truth across all
platforms.
%% ====================================================================
\section{Summary and Outlook}
\label{sec:summary}
%% ====================================================================

We have introduced a scalable, planted-solution family of SAT and Ising benchmark instances derived from integer factorization. The construction maps the arithmetic constraints of binary multiplication into a constraint system whose satisfying assignments correspond to valid factorizations of $N = pq$, with the known pair $(p,q)$ serving as a built-in ground truth.

We found that for $d$-bit primes, the $d^4$ growth of variables and clauses arises from carry cascading: each contraction generates carries that propagate across columns, producing a quadratic growth in column population and, upon summation, a quartic total. This mechanism induces long-range correlations across the instance, qualitatively distinguishing these benchmarks from random $k$-SAT.

Empirical benchmarks with state-of-the-art SAT solvers show that median runtime is consistent with exponential scaling in the bit-length~$d$, roughly doubling with each additional bit over the range tested. While integer factorization is widely believed to be computationally hard, this does not automatically translate to hardness of a particular SAT encoding; nevertheless, the observed scaling indicates that the present construction produces instances of rapidly increasing difficulty. Whether this trend persists asymptotically remains an open question.

From a practical perspective, the construction provides a simple, deterministic pipeline that generates instances controlled by a single parameter $d$, requiring only the ability to sample $d$-bit primes. The resulting benchmarks are directly usable across multiple platforms: SAT solvers (via DIMACS CNF), classical optimization algorithms (via Ising compilation), and quantum optimization devices after embedding.

Overall, this benchmark family occupies a complementary regime to existing generators: it combines exact analytical scaling, rich arithmetic structure, planted-solution verifiability, and dual CNF/Ising representations. As such, we believe it provides a controlled setting for probing solver behavior on structured, non-random instances with intrinsic long-range correlations. Open-source software implementing the full pipeline is available
at~\cite{software}.

%% ====================================================================
\begin{acknowledgments}
We acknowledge the Center for Advanced Research Computing (CARC) at the University of Southern California for providing computing resources that have contributed to the research results reported in this study (\url{https://carc.usc.edu}). The author used AI tools for stylistic and grammatical editing and as a programming assistant for code generation. All algorithms and specifications were designed by the author. No AI was used to generate or modify the mathematical results, proofs, numerical experiments, or scientific conclusions. All code was reviewed and validated by the author.
\end{acknowledgments}
%% ====================================================================

%% ====================================================================
\bibliography{refs}
%% ====================================================================

\appendix

%% ====================================================================
\section{Worked Example: $p = 11$, $q = 13$}
\label{app:worked_example}
%% ====================================================================

This appendix walks through the full pipeline for
$N = 143 = 11 \times 13$.  Because both factors are $4$-bit integers,
this is the smallest instance of the symmetric case $n_p = n_q = d$
with $d = 4$, making the example a direct illustration of the
scaling formulas of Sec.~\ref{sec:scaling}.

%% ------------------------------------------------------------------
\subsection{Binary representations}

The two primes and their product in least-significant-bit-first
(LSB-first) binary are
\begin{align}
  p &= 11 = (1\;1\;0\;1)_2, &  p_0 &= 1,\; p_1 = 1,\; \nonumber\\
  &&p_2& = 0,\; p_3 = 1, \nonumber \\[2pt]
  q &= 13 = (1\;0\;1\;1)_2, & q_0 &= 1,\; q_1 = 0,\; \nonumber\\
  &&q_2& = 1,\; q_3 = 1,
  \label{eq:app_pq} \\[2pt]
  N &= 143 = (1\;1\;1\;1\;0\;0\;0\;1)_2, & N_0 &= 1,\; N_1 = 1,\; N_2 = 1,\; \nonumber\\
  && N_3 & = 1,  N_4 = 0,\; \nonumber\\
  &&N_5 &= 0,\; N_6 = 0,\; N_7 = 1.  \nonumber
\end{align}
Here $n_p = n_q = d = 4$.

%% ------------------------------------------------------------------
\subsection{Partial products}

The $4 \times 4 = 16$ partial products $a_{ij} = p_i \wedge q_j$
form the multiplication table
\begin{equation}
\begin{array}{c|cccc}
  & q_0{=}1 & q_1{=}0 & q_2{=}1 & q_3{=}1 \\[2pt]
\hline
\rule{0pt}{10pt}
p_0{=}1 & a_{00}{=}1 & a_{01}{=}0 & a_{02}{=}1 & a_{03}{=}1 \\
p_1{=}1 & a_{10}{=}1 & a_{11}{=}0 & a_{12}{=}1 & a_{13}{=}1 \\
p_2{=}0 & a_{20}{=}0 & a_{21}{=}0 & a_{22}{=}0 & a_{23}{=}0 \\
p_3{=}1 & a_{30}{=}1 & a_{31}{=}0 & a_{32}{=}1 & a_{33}{=}1 \\
\end{array}
\label{eq:app_pp_matrix}
\end{equation}
where each $a_{ij}$ is placed in column $k = i + j$ of the
addition table.  The column assignment is shown in
Table~\ref{tab:app_pp}.

\begin{table}[h]
\centering
\caption{Partial products assigned to each column.}
\label{tab:app_pp}
\small
\begin{tabular}{cll}
\toprule
Column $k$ & Partial products & $\mathrm{pp}_k$ \\
\midrule
0 & $a_{00}$ & 1 \\
1 & $a_{01},\; a_{10}$ & 2 \\
2 & $a_{02},\; a_{11},\; a_{20}$ & 3 \\
3 & $a_{03},\; a_{12},\; a_{21},\; a_{30}$ & 4 \\
4 & $a_{13},\; a_{22},\; a_{31}$ & 3 \\
5 & $a_{23},\; a_{32}$ & 2 \\
6 & $a_{33}$ & 1 \\
\bottomrule
\end{tabular}
\end{table}

%% ------------------------------------------------------------------
\subsection{Column contraction}

The contractions proceed left-to-right.  At column~$k$, the
column contains its $\mathrm{pp}_k$ partial products plus carries
arriving from column~$k{-}1$; if it holds $m_k \geq 2$ entries,
$m_k - 1$ half-adder contractions reduce it to a single entry, each
generating one XOR clause (sum) and one AND clause (carry), with the
carry promoted to column~$k{+}1$.

Table~\ref{tab:app_contraction} lists $m_k$ and the number of
contractions at each column.  As predicted by
Eq.~\eqref{eq:mk_asc}, the ascending phase ($k = 0, \ldots, 3$)
gives $m_k = 1 + k(k{+}1)/2$, i.e., $m_0 = 1$, $m_1 = 2$,
$m_2 = 4$, $m_3 = 7$.  The peak occurs at $k = 2d{-}2 = 6$:
\begin{equation*}
m_6 = d^2 - 2d + 2 = 16 - 8 + 2 = 10\,,
\end{equation*}
matching Eq.~\eqref{eq:peak}.  The tail phase
($k \geq 2d{-}1 = 7$) decreases by one per column until $m_{15} =
1$.  The last active column is $k_{\max} = d^2 - 1 = 15$.

The total number of contractions is
\begin{equation*}
C(4) = \tfrac{4^2 \cdot 3^2}{2} = 72\,,
\end{equation*}
consistent with Eq.~\eqref{eq:Cd} and the column-by-column sum in
Table~\ref{tab:app_contraction}.

\begin{table}[h]
\centering
\caption{Column contraction for $N = 143$.  Each contraction
produces one XOR clause and one carry (AND clause).  Columns $k =
0, \ldots, 6$ receive partial products; columns $k \geq 7$ contain
only carries.}
\label{tab:app_contraction}
\small
\begin{tabular}{crrrrc}
\toprule
Column $k$ & $m_k$ & Contractions & XOR & Carry & $N_k$ \\
\midrule
0  &  1  &  0  &  0  &  0  &  1  \\
1  &  2  &  1  &  1  &  1  &  1  \\
2  &  4  &  3  &  3  &  3  &  1  \\
3  &  7  &  6  &  6  &  6  &  1  \\
4  &  9  &  8  &  8  &  8  &  0  \\
5  &  10  &  9  &  9  &  9  &  0  \\
6  &  10  &  9  &  9  &  9  &  0  \\
7  &  9  &  8  &  8  &  8  &  1  \\
8  &  8  &  7  &  7  &  7  &  0  \\
9  &  7  &  6  &  6  &  6  &  0  \\
10 &  6  &  5  &  5  &  5  &  0  \\
11 &  5  &  4  &  4  &  4  &  0  \\
12 &  4  &  3  &  3  &  3  &  0  \\
13 &  3  &  2  &  2  &  2  &  0  \\
14 &  2  &  1  &  1  &  1  &  0  \\
15 &  1  &  0  &  0  &  0  &  0  \\
\midrule
\multicolumn{2}{c}{Total} &  72  &  72  &  72 \\
\bottomrule
\end{tabular}
\end{table}

\begin{figure*}[t]
\centering
\includegraphics[width=\textwidth]{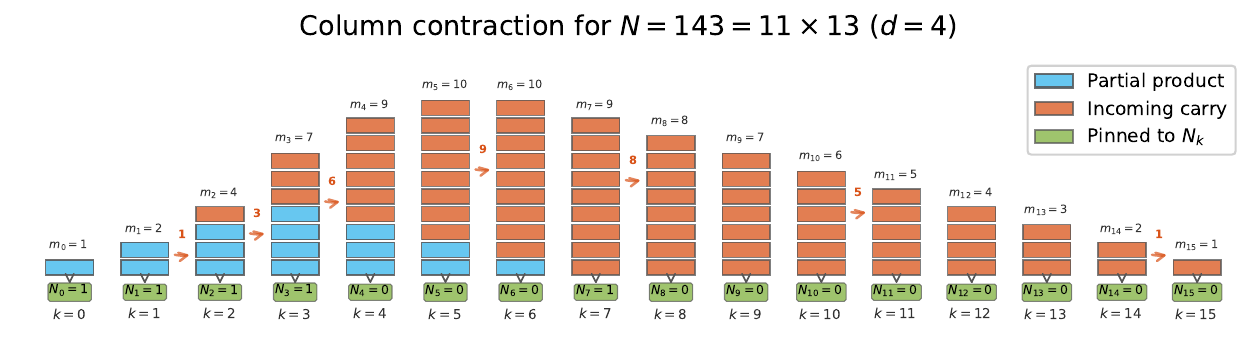}
\caption{Schematic of the column contraction for
$N = 143 = 11 \times 13$.  Each column~$k$ contains $m_k$ entries:
partial products from the multiplication table (blue) and carries
arriving from column~$k{-}1$ (orange).  Arrows indicate carry
propagation with the number of carries labeled.  After contraction,
the single remaining entry in each column is pinned to the known
bit $N_k$ (green).  The column heights directly reflect the $m_k$
profile of Fig.~\ref{fig:mk_profiles}(a) for $d = 4$.}
\label{fig:carry_flow}
\end{figure*}

Figure~\ref{fig:carry_flow} provides a schematic view of the
contraction process.  The column heights directly visualize the
$m_k$ sequence from Table~\ref{tab:app_contraction}: partial
products (blue) seed the low-order columns, while incoming carries
(orange) increasingly dominate at higher columns, driving the
population toward its peak at $k = 5$--$6$.  The pinned output
$N_k$ at the base of each column shows the known bits of~$N$ that
anchor the constraint propagation.

%% ------------------------------------------------------------------
\subsection{Pre-reduction instance sizes}

The pre-reduction system contains:

\begin{itemize}
\item \textbf{Boolean variables}: $2d + d^2 + 2C(d) = 8 + 16 +
  144 = 168$.  These comprise the $2d = 8$ input bits ($p_0,
  \ldots, p_3, q_0, \ldots, q_3$), $d^2 = 16$ partial products,
  and $2 \times 72 = 144$ auxiliary variables (one sum and one
  carry per contraction).
\item \textbf{AND clauses}: $d^2 + C(d) = 16 + 72 = 88$ (16 from
  partial products, 72 from carries).
\item \textbf{XOR clauses}: $C(d) = 72$.
\item \textbf{Pinning constraints}: $d^2 = 16$ (one per active
  column).
\item \textbf{Total constraints}: $2d^2 + d^2(d{-}1)^2 = 32 + 144 = 176$.
\end{itemize}

\noindent All of these match the formulas in Table~\ref{tab:formulas}
exactly.

%% ------------------------------------------------------------------
\subsection{Preprocessing}
\label{subsec:app_preprocessing}

The iterative Boolean reduction proceeds as follows.  We describe
the key deductions grouped by the column structure that generates
them.

\paragraph{Column 0: extracting the LSBs.}
The sole entry in column~0 is $a_{00} = p_0 \wedge q_0$, pinned to
$N_0 = 1$.  Since an AND output of~1 forces both inputs to~1, we
immediately deduce
\begin{equation*}
p_0 = 1, \qquad q_0 = 1.
\end{equation*}
This collapses all partial products involving $p_0$ or $q_0$:
\begin{equation*}
a_{0j} = 1 \wedge q_j = q_j, \qquad
a_{i0} = p_i \wedge 1 = p_i,
\end{equation*}
for all $i, j$.  In particular, $a_{01} = q_1$, $a_{02} = q_2$,
$a_{03} = q_3$, $a_{10} = p_1$, $a_{20} = p_2$, and $a_{30} = p_3$,
reducing these six AND clauses to equivalences.

\paragraph{Column 1: linking $p_1$ and $q_1$.}
After simplification, column~1 contains two entries: $a_{01} = q_1$
and $a_{10} = p_1$.  Their XOR is pinned to $N_1 = 1$:
\begin{equation*}
q_1 \oplus p_1 = 1 \qquad\Longrightarrow\qquad
p_1 = \lnot\, q_1.
\end{equation*}
The carry from this column is $q_1 \wedge p_1 = q_1 \wedge \lnot\,
q_1 = 0$.  This zero carry propagates into column~2, effectively
removing one entry from it.

\paragraph{Column 2: linking $p_2$ and $q_2$.}
Column~2 receives partial products $a_{02} = q_2$, $a_{11} = p_1
\wedge q_1$, $a_{20} = p_2$, plus the zero carry from column~1.
The substitution $p_1 = \lnot\, q_1$ gives $a_{11} = \lnot\, q_1
\wedge q_1 = 0$, so two of the four entries vanish.  The surviving
entries are $q_2$ and $p_2$; a XOR contraction eventually pins
through to $N_2 = 1$, yielding
\begin{equation*}
p_2 = \lnot\, q_2
\end{equation*}
(after propagation through the carry chain).

\paragraph{Subsequent iterations.}
Each further preprocessing iteration propagates carry-chain
simplifications deeper into the column structure.  The zero products
arising from $p_2 = 0$ (in the planted solution) cascade through
columns~2--5, collapsing additional AND clauses.  The complementarity
relations $p_1 = \lnot\, q_1$ and $p_2 = \lnot\, q_2$ further
reduce XOR clauses that share these variables.

Analogously, the high-column pins ($N_k = 0$ for $k = 8, \ldots,
15$) propagate backward: a column pinned to~0 whose carry chain
produces only zero carries forces its entries to specific values,
which in turn constrain the partial products and hence the input
bits.

\paragraph{Convergence.}
After convergence, the preprocessing has resolved 139 of the
168~variables: 77~are pinned to fixed values and 62~are merged
into equivalence classes via the union-find structure (identified
as equal or complementary to other variables).  This leaves
29~independent free Boolean variables.  The residual system
consists of 20~AND and 12~XOR clauses, converting to 102~CNF
clauses---a reduction of over 80\%\ from the 552 raw CNF clauses
that the unpreprocessed instance would contain.

Of the eight input bits, $p_0$ and $q_0$ are pinned to~1.  The
remaining six are free but linked pairwise by the equivalences
\begin{equation}
p_1 \equiv \lnot\, q_1, \qquad
p_2 \equiv \lnot\, q_2, \qquad
p_3 \equiv a_{30}, \qquad
q_3 \equiv a_{03},
\label{eq:app_equivs}
\end{equation}
where $a_{30}$ and $a_{03}$ are partial-product variables that
have been identified with $p_3$ and $q_3$, respectively, by the
reduction $q_0 = p_0 = 1$.  These equivalences reduce the
effective degrees of freedom: the SAT solver need only determine
the values of $q_1$, $q_2$, $q_3$ (and the internal auxiliary
variables) to reconstruct the full factorization.

%% ------------------------------------------------------------------
\subsection{CNF output}

The 20~residual AND clauses each produce 3~CNF clauses
[Eq.~\eqref{eq:and_cnf}], and the 12~residual XOR clauses each
produce 4~CNF clauses [Eq.~\eqref{eq:xor_cnf}], giving a raw
count of $3 \times 20 + 4 \times 12 = 108$ clauses.  After
removing tautologies and subsumed clauses during the
CNF cleanup pass, 102~clauses survive on 29~variables.

%% ------------------------------------------------------------------
\subsection{Ising compilation}

The Ising model is assembled from the 20~residual AND clauses and
12~residual XOR clauses.  Each AND gadget
[Eq.~\eqref{eq:and_gadget}] contributes 3 couplings with no
auxiliary spins; each XOR gadget [Eq.~\eqref{eq:xor_gadget}]
contributes 6~couplings and introduces one auxiliary spin.  The
resulting Hamiltonian has $29 + 12 = 41$ spins (29~physical, 12
XOR auxiliaries) before coupling merging.

After merging duplicate spin pairs, the Ising model contains 25
nonzero local fields and 116 nonzero couplings, with a constant
offset $E_0 = 107$.  The planted spin configuration achieves
$H(\mathbf{s}^{\star}) = E_0$ (verified numerically), and the
minimum excitation gap is $\Delta = 2$.

%% ------------------------------------------------------------------
\subsection{Summary of instance dimensions}

Table~\ref{tab:app_summary} collects all key quantities for this
example, comparing the pre-reduction formula predictions with the
actual values.

\begin{table}[hbp]
\centering
\caption{Instance dimensions for $N = 143 = 11 \times 13$
($d = 4$).  Pre-reduction values match the exact formulas of
Table~\ref{tab:formulas}.}
\label{tab:app_summary}
\small
\begin{tabular}{lrr}
\toprule
Quantity & Pre-reduction & Post-reduction \\
\midrule
Boolean variables     & 168  & 29 (free) \\
Variables pinned      & ---  & 77 \\
Variables merged      & ---  & 62 \\
AND clauses           & 88   & 20 \\
XOR clauses           & 72   & 12 \\
Pinning constraints   & 16   & --- \\[3pt]
CNF variables         & ---  & 29 \\
CNF clauses           & 552$^{\,\dagger}$ & 102 \\[3pt]
Ising spins           & ---  & 41 \\
Ising couplings       & ---  & 116 \\
\bottomrule
\multicolumn{3}{l}{\footnotesize ${}^{\dagger}$\,Without preprocessing:
$3 \times 88 + 4 \times 72 = 552$ raw;}\\
\multicolumn{3}{l}{\footnotesize \phantom{${}^{\dagger}$\,}511 after
pin substitution and clause cleanup.}
\end{tabular}
\end{table}

\end{document}